\documentclass[journal, 12pt, draftclsnofoot, one column]{IEEEtran}
\IEEEoverridecommandlockouts
\usepackage{amsmath,amssymb,amsfonts}
\usepackage{algorithmic}
\usepackage{graphicx}
\graphicspath{{Figures/}}

\usepackage{amsthm}

\newtheorem{definition}{Definition}

\usepackage{enumitem}

\usepackage{textcomp}
\usepackage{xcolor}
\def\BibTeX{{\rm B\kern-.05em{\sc i\kern-.025em b}\kern-.08em
    T\kern-.1667em\lower.7ex\hbox{E}\kern-.125emX}}
\begin{document}

\title{\huge Autonomous Mobility and Energy Service Management in Future Smart Cities: An Overview}

%\author{Last Modified on \today\\
%	Drafted by Xiaoqi Tan, ECE, University of Toronto, Canada}

\author{\IEEEauthorblockN{Xiaoqi Tan, Alberto Leon-Garcia}
\thanks{The authors are the Edward S. Rogers Sr. Department of Electrical \& Computer Engineering , University of Toronto. Email: \{xiaoqi.tan, alberto.leongarcia\}@utoronto.ca.}
%\IEEEauthorblockA{Electrical and Computer Engineering\\University of Toronto \\xiaoqi.tan@utoronto.ca} \and

%\IEEEauthorblockN{Alberto Leon-Garcia}
%\IEEEauthorblockA{Electrical and Computer Engineering\\ University of Toronto 	alberto.leongarcia@utoronto.ca}

%\and 

%\IEEEauthorblockN{Danny H.K. Tsang}
%	\IEEEauthorblockA{Electronic and Computer Engineering\\Hong Kong University of Sci. and Tech. \\
%		eetsang@ust.hk}
	
%\and

}

\maketitle

\begin{abstract}
With the rise of transportation electrification, autonomous driving and shared mobility in urban mobility systems,  and increasing penetrations of distributed energy resources and autonomous demand-side management techniques in energy systems, tremendous opportunities, as well as challenges, are emerging in the forging of a sustainable and converged urban mobility and energy future. This paper is motivated by these disruptive transformations and gives an overview of managing autonomous mobility and energy services in future smart cities. First, we propose a three-layer architecture for the convergence of future mobility and energy systems.  For each layer, we give a brief overview of the disruptive transformations that directly contribute to the rise of autonomous mobility-on-demand (AMoD) systems. Second, we propose the concept of autonomous flexibility-on-demand (AFoD), as an energy service platform built directly on existing infrastructures of AMoD systems. In the vision of AFoD, autonomous electric vehicles provide charging flexibilities as a service on demand in energy systems. Third, we analyze and compare AMoD and AFoD, and we identify four key decisions that, if appropriately coordinated, will create a synergy between AMoD and AFoD. Finally, we discuss key challenges towards the success of AMoD and AFoD in future smart cities and present some key research directions regarding the system-wide coordination between AMoD and AFoD.
\end{abstract}


\section{Introduction}
It is estimated that over 50\% of world's population is now living, commuting and working in urban areas and that the number will rise to 70\% by 2050 \cite{Bain&Company2018}. The acceleration of urbanization is also evidenced by the significant expansion of major cities in the last several decades \cite{LeannaGarfield}. This has created enormous problems with respect to environmental pollution and the
general quality of life of urban residents.  According to \cite{Opec2015}, road transportation contributes over 40\% of global oil consumption and over 70\% of greenhouse gas emission from the sector. Worse still, as city population and urban areas continue to grow,  the average commuting distance increases,  and thus more traffic will be generated due to the increased demand for personal mobility. As a consequence, air pollution, traffic congestion, and limited parking spaces are becoming severe urban problems confronting residents, government and other related urban stakeholders. 
%In particular, the so-called ``big city disease" in China has 

%In addition to the challenges coming from urban mobility systems, energy systems also play a pivotal role in shaping future urban sustainability. 

Motivated by the aforementioned problems, current mobility systems are undergoing dramatic transformations.  First, electric vehicles (EVs) have much higher energy efficiency and zero emission capability compared to their gasoline counterparts. Therefore, \textit{transportation electrification} is viewed as a promising solution to tackle oil dependency and climate change \cite{M.Alexander2016}. Second, privately-owned vehicles are typically parked more than 90\% of the time, and thus it is widely believed that the traditional vehicle ownership model is unsustainable for future personal mobility \cite{Mitchell2013}.  \textit{Mobility-on-Demand (MoD)} is emerging as the revolutionary business model for personal mobility with a user-centric, smartphone-based on-demand system. According to \cite{Shaheen2017}, MoD is defined as an innovative transportation concept where consumers can access mobility, goods, and services on demand by dispatching or using shared mobility.  Augmented with transportation electrification and the rapidly-developing \textit{autonomous driving technology}, autonomous MoD, or AMoD in short, has the potential to fundamentally change the future of automobile industry and the way we commute in cities \cite{Mitchell2013}. 

Disruptive transformations in energy systems and urban mobility systems form the twin pillars for creating urban sustainability.  Specifically, the production of electricity is very different today than before. For example, in the past several decades, the U.S. energy supply had a significant shift from wood, coal, and oil towards natural gas and now renewables \cite{energyChanges}. With ever-increasing intermittent, small, and grid-connected storage and renewable energy generators that are typically referred to as distributed energy resources (DERs), e.g., rooftop and community solar and wind turbines, our ability to control the electricity generation to meet demand in real-time is being challenged. Therefore, a more flexible system that can quickly adapt to the intermittency of \textit{high penetrations of DERs} is of paramount importance. \textit{Autonomous demand-side management (DSM)} deals with techniques that can automatically modify consumer demand based on real-time bidirectional digital communication between generators and loads, and is a promising solution for future energy systems with high penetrations of DERs \cite{flexibility_on_demand}. 

%Over the past decade, significant achievements have been reported from both the academia and industry on the development of autonomous DSM programs \cite{smart_grid}.

%As such that the flexibility in the demand-side can be fully elicited and exploited. 

%Interfaced by autonomous EVs, the future of urban mobility systems and energy systems is converging, which creates a substantial new opportunities as well as challenges. 

%Facing all the aforementioned issues, mobility and energy systems in cities will have to undergo significant transformations to become sustainable and affordable with customer-centric infrastructure and services. There is an ongoing trend in transportation domain in the following three aspects.

%\subsection{Transformation of Mobility Systems}

%In particular, we focus on the hybrid service management of urban mobility and energy systems interfaced by a network of AEVs.  

%Motivated by the challenges and rapid changes in urban mobility and energy systems, t

%\footnote{However, this paper is not intended to be a comprehensive review of the state of the art, but mainly consists of some interesting and promising directions that are important in achieving urban sustainability.}

Motivated by the above disruptive transformations in urban mobility and energy systems, this paper aims to give an overview of two promising autonomous services in future smart cities. \textit{First}, we propose a three-layer architecture to show the convergence of mobility and energy systems in future smart cities.  For each layer, we give a brief overview of the disruptive transformations that directly contribute to the rise of AMoD systems. \textit{Second}, we propose the concept of autonomous flexibility-on-demand (AFoD), which is a promising energy service platform built directly on existing infrastructures of AMoD systems. In the vision of AFoD, autonomous EVs in AMoD systems leverage their charging flexibilities to provide autonomous charging/discharging services on demand. \textit{Third}, we perform a conceptual comparison between AMoD and AFoD and identify four key decisions that, if appropriately coordinated, will create a synergy between AMoD and AFoD. \textit{Finally}, we summarize some key challenges towards the success of AMoD and AFoD in future smart cities and present some key research directions regarding the system-wide coordination between AMoD and AFoD.

%In particular, we focus on AMoD systems by first proposing a three-layer architecture to show the convergence of urban mobility and energy systems. Second, we propose a new concept called autonomous flexibility-on-demand (AFoD), which is believed to be a promising low-cost solution to enable future energy systems to quickly adapt to the intermittency originated from DERs. The beauty of AFoD is that it leverages the existing infrastructures of AMoD systems, and a promising synergy effect between AMoD and AFoD can be achieved if there exist appropriate system-wide coordination and management methods. Finally, we briefly review the key challenges and identify some open questions regarding the future of AMoD and AFoD. 

The rest of this paper is organized as follows.  We give an overview of AMoD systems in Section \ref{AMoD_overview}, which includes a layered-view of architectures, key disruptive transformations of each layer, and the potential of AMoD systems. Section \ref{AFoD_overview} introduces the concept of AFoD and provides a detailed discussion on the comparison and synergy between AMoD and AFoD. Section \ref{Challenges} summarizes the key challenges that determine the future of AMoD and AFoD. We conclude the paper in Section \ref{Conclusions}. 

\section{AMoD Systems: Architecture, \\Disruptive Transformations and Potential} \label{AMoD_overview}
In this section, we give an overview of AMoD systems and some key transformations towards the convergence of urban mobility and energy systems. We also discuss the limitations of current MoD systems and anticipate the potential of AMoD systems in creating urban sustainability.

\begin{figure*}
	\centering
	\includegraphics[width=16cm]{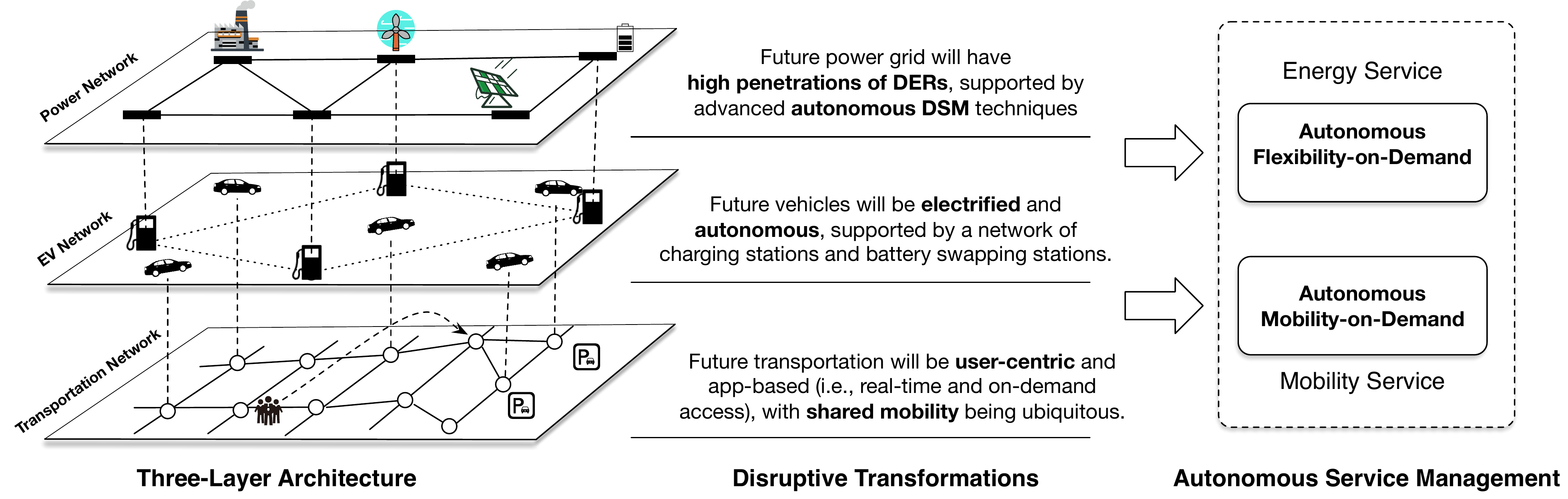}
	\caption{Left: the three-layer architecture of urban mobility and energy systems. Middle: key disruptive transformations of each layer.  Right: two promising autonomous service platforms for future mobility and energy systems. }
	\label{AMoD}
\end{figure*}

\subsection{Architectural Overview} \label{overview_architecture}
With the rising popularity of battery-powered EVs, the transportation sector is becoming increasingly influenced by the electricity supply from the power sector. Meanwhile, the spatial-temporal uncertain charging loads from EVs also yield considerable challenge to the infrastructures of current power networks \cite{M.Alexander2016}.  To illustrate this mutual relationship, we provide a layered-architecture  in the left side of Fig. \ref{AMoD}. 

Specifically, the \textit{power network} sits at the upper layer, linked with the middle layer through geographically-distributed charging facilities. The charging facilities will be connected to different nodes/buses of the power distribution networks.  In the middle layer, there is an \textit{EV network}, which serves as an interface between the upper-layer power network and the bottom-layer \textit{transportation network}. Note that an EV network in our context is defined as a network of EVs and the related charging facilities.   In the bottom layer, the transportation networks provide all the necessary transportation facilities to the EV network such as road networks, traffic lights, and parking spaces, etc.

%We give an brief overview of the key techniques that play a critical role in creating urban sustainability, namely i) the integration of DERs and demand-side management in power networks, ii) the autonomous driving techniques and energy refueling of EVs in electric vehicle networks, and iii) the user-centric and sharing-based mobility in transportation networks. These three aspects correspond to the three layers of urban mobility and energy systems depicted in Fig. \ref{AMoD}. 

\subsection{Disruptive Transformations of Each Layer}
\label{disruptive_transformation}
As shown in the middle of Fig. \ref{AMoD}, each layer of the three-layer architecture is undergoing some disruptive transformations that, if fully realized, will fundamentally reshape the future of urban mobility and energy systems.  Below we provide a brief overview of these transformations. 
%Note that our overview here is not intended to be comprehensive, but mainly centers on the driving forces of these transformations as well as their overall impact to the development of AMoD systems.

\subsubsection{\textbf{Upper Layer (Power Networks)}} The current power grid is undergoing a dramatic change in order to meet the economic and environmental sustainability requirement. In particular,  \textit{high penetrations of DERs} and development of \textit{autonomous DSM} techniques are two critical transformations. 

\begin{itemize}[leftmargin=*]
\item \textit{High Penetrations of DERs}. The current power grid is expected to evolve significantly over the next few decades to become more sustainable and efficient. One of the most significant evolutions is the idea of integrating DERs at the edge of the power grid \cite{Bain&Company2018} \cite{InternationalEnergyAgency2011}.  Specifically, DERs are a group of small-scale, grid-connected storage and renewable energy generation resources that are capable of injecting power into the power grid. Typical examples of DERs include rooftop and community solar, wind turbines, combined heat power, and microgrids, etc. DERs are usually modularized and flexible, and thus they can be easily purchased and installed by end-users without large investment on extra power transmission lines. Meanwhile, DERs can reduce the total amount of power transmission loss since the electricity generation is close to where it is used.

%However, In the absence of effective tools and methodologies to manage high levels of DERs, the power grid will experience frequency and voltage variations, overloads of transformers and transmission lines, phase load imbalances, and other variations from  operating standards of power grids. Given these challenges, new tools and methodologies must be developed for the technical and economic management of power grids with high penetration of DERs.  

\item \textit{Autonomous DSM}. One of the central issues confronting current grid operators is that existing power distribution systems are unaccustomed to intermittency, a major characteristic of DERs \cite{InternationalEnergyAgency2011}.  In the absence of effective tools and methodologies to manage high levels of DERs, the power grid will experience frequency and voltage variations, overloads of transformers and transmission lines, phase load imbalances, and other variations from  operating standards of power grids. Given these challenges, new tools and methodologies must be developed for the technical and economic management of power grids with high penetrations of DERs. Facilitated by the bidirectional communication between electricity generation and consumption, autonomous DSM  is a  promising solution to these problems. Over the past decade, significant achievements have been reported from both academia and industry on the development of autonomous DSM programs to help support the integration of DERs, e.g.,  \cite{flexibility_on_demand}-\cite{Mohsenian-Rad2010}.

\end{itemize} 

\subsubsection{\textbf{Middle Layer (EV Networks)}} Key technologies driving the transformations of the middle layer lie at \textit{transportation electrification} and \textit{autonomous driving}. 
 
\begin{figure}
	\centering
	\includegraphics[width=12cm]{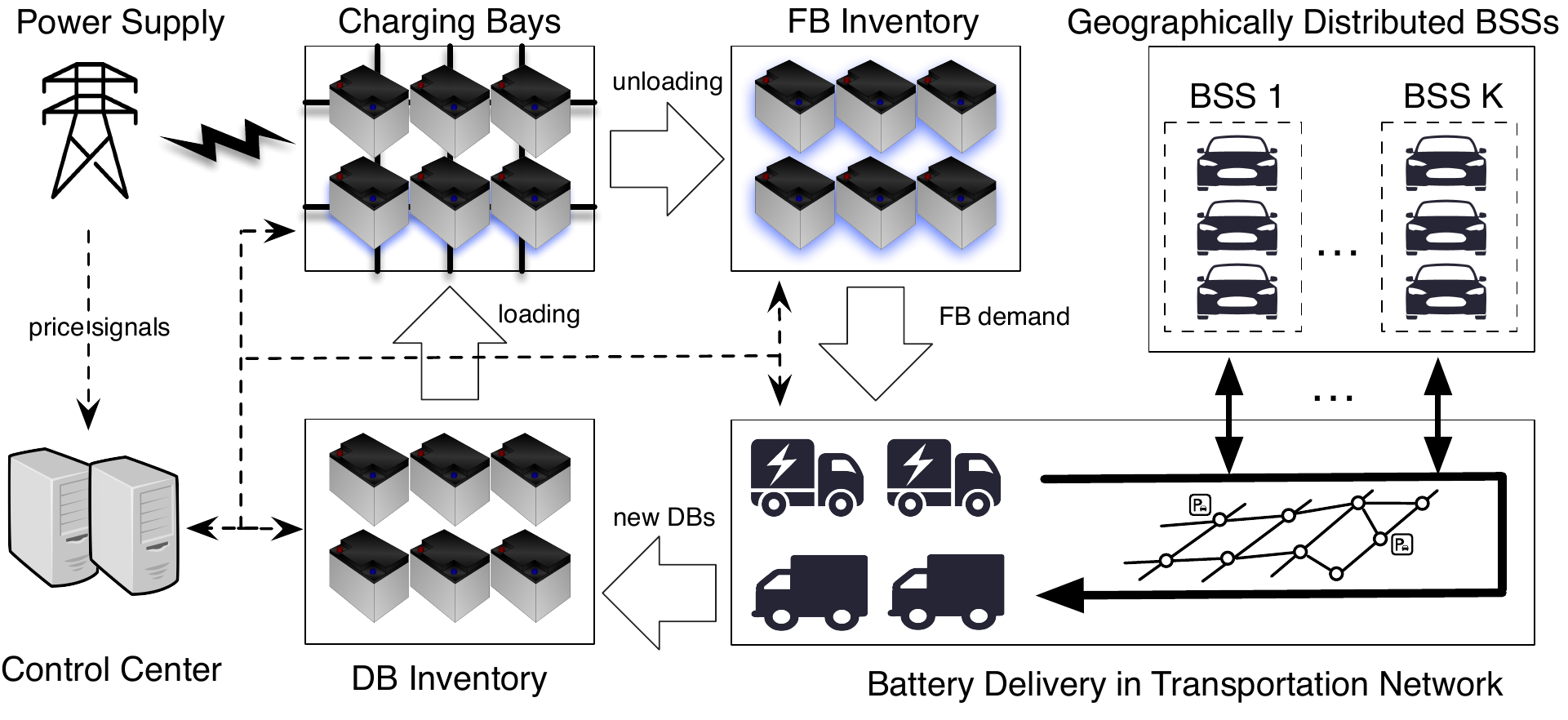}
	\caption{A candidate approach to implement distributed battery swapping stations and centralized charging stations in EV networks.  Suppose there exist multiple geographically distributed battery swapping stations (BSSs) in an urban area. Depleted batteries (DBs) are unloaded from vehicles and then collectively delivered to a DB inventory where battery charging service is provided. After being fully-charged, those fully-charged batteries (FBs) will be stored in FB inventory and delivered back to BSSs based on their different local demand. More detailed information is referred to \cite{own}. }
	\label{BSS}
\end{figure}

\begin{itemize}[leftmargin=*]
\item \textit{Transportation Electrification}. Driven by reducing oil dependency and the impact of vehicles on the environment, many countries and cities have announced goals to eventually ban internal combustion engines \cite{PaulA.Eisenstein}. 
Key to the success of transportation electrification lies in battery technologies as well as the deployment of charging facilities. As suggested by \cite{Bain&Company2018}, currently the major obstacles preventing a wider adoption of EVs is the price of batteries and the long charging time. However, given the active research and development of technologies and business models of batteries and charging facilities all over the world, e.g., leading automobile and charging facility manufacturers such as Tesla \cite{Tesla},  it is anticipated that the transportation electrification will keep accelerating in the following decades \cite{Bain&Company2018, M.Alexander2016}. Moreover, an increasing interest in battery swapping models has developed in recent years, mainly focused on public EVs such as electric taxis and buses, etc. For instance,  in India, Sun Mobility is developing a service for swapping electric bus batteries, as well as smaller two- and three-wheel vehicles \cite{sun_mobility}. In China, the province of Zhejiang is developing a network of  battery swapping and charging stations \cite{Bain&Company2018} serving its electric taxi networks, whose operational process is illustrated in Fig. \ref{BSS}.

%Fig. \ref{BSS} shows an example of how to implement distributed battery swapping stations and centralized charging stations in an urban area. 

\item  \textit{Autonomous Driving}. An autonomous vehicle is a vehicle that can maneuver without human intervention. As the core technology in the development of AMoD systems, autonomous driving technology will disrupt the status quo and fundamentally reshape urban life
and the global economy. It is, therefore, considered to be one of the most highly anticipated technologies over the coming decade, as McKinsey Global Institute reported in \cite{disruptive_technologies}. Direct benefits of being driverless include i) increased safety due to the removal of human errors, ii) increased convenience and productivity as humans are freed from the burden of driving, and iii) reduced impact on the environment as velocity can be precisely tuned by the computer to minimize emission and noise. Driven by these significant benefits and the huge potential market \cite{disruptive_technologies}, a number of companies have invested in the research and development of autonomous driving and its related business models, e.g., ranging from Waymo \cite{Waymo} (former Google Self-Driving Car Project), to the General Motors EN-V project \cite{GM} centered on the operation and management of autonomous urban mobility systems.  
\end{itemize}

\subsubsection{\textbf{Bottom Layer (Transportation Networks)}}
The transformation of the bottom layer is driven by the innovative and disruptive business models,  as well as technological innovations in sensing, monitoring, and controlling the transportation network and the development of vehicle connectivity, as evident in the rise of MoD such as car-sharing, ride-sharing, and bikesharing, etc. In reality, recent years have witnessed a worldwide success of MoD platforms such as Uber and Lyft, proving that this highly user-centric and shared mobility based transportation service is a promising combination of economic, environmental, and social forces \cite{Mitchell2013, Shaheen2017}. In 2016, the shared mobility market in China, Europe and the United States was nearly \$54 billion \cite{shared_mobility_market}. Under the most positive scenario,  as anticipated by 
McKinsey, the market for the shared mobility could have 28 percent  annual growth from 2015 to 2030 \cite{shared_mobility_market}.  

The growth of MoD systems has a close tie to the increasing popularity of sharing economy in many economic sectors, where the biggest conceptual change lies in receiving goods and services based on sharing, renting and borrowing, rather than owning them \cite{Shaheen2017}. 
%MoD is such an innovative and disruptive transportation concept where consumers can access mobility services on demand by using shared mobility.  
From the technology perspective, many factors contribute to the rapid growth of MoD systems in practice. For instance, advancement in communication and cloud computing technologies enables MoD customers to have real-time access to  mobility services. Moreover, the increasing popularity of peer-to-peer and mobile transactions makes pricing and payment in MoD systems easy and convenient. For instance, the convenience of mobile payments in China (e.g., Alipay and Wechat Pay) significantly contribute to the rapid success of DiDi, the biggest ride-sharing service platform in China and also one of the biggest global competitors to Uber.   

%In summary, 

%Examples of MoD systems include car-sharing, ride-sharing and bikesharing, etc. We give an brief overview of the first two systems as follows:
%\begin{itemize}
%	\item \textit{Car-sharing}. It is widely known that privately-owned vehicles are typically parked more than 90\% of the time. Therefore, \textit{one-way car-sharing} is believed to be a promising strategy. 
%	
%	\item \textit{Ride-sharing}. Increased urbanization and changing demographics. Reduce oil dependency, pollution, promote higher utilization rates and reduce parking lot sprawls.
%\end{itemize}

%\footnote{The major focus of this paper is on autonomous EVs, thus bikesharing is omitted in our discussion.}

%The disruptive technologies and business models in each layer, if fully realized, show a clear trend of convergence of urban mobility and energy systems. 

\subsection{Beyond MoD: AMoD and Its Potential}\label{Autonmous_MoD}
%This section introduces the limitations of current MoD systems and anticipate the rise of AMoD systems in creating  a sustainable urban mobility system.
%However, the power of AMoD systems is substantially superior to traditional MoD systems.

%\subsubsection{Limitations of MoD Systems}
Despite the huge economic success of existing MoD platforms, MoD systems, however, have some limitations due to the nature of shared mobility. First, vehicles will automatically concentrate in  certain areas since the customers are not evenly distributed in the spatial-temporal domain. Second, car-sharing does not directly contribute to congestion reduction, as at least the same amount of distance would be traveled with the same amount of origin-destination requests, or even more when considering the extra traveling distance due to balancing. The same problem also exists in the case of ride-sharing. For instance, a study for the Boston area shows that the app-based ride-sharing services have indeed generated more traffic on city streets \cite{Gehrke2018}. 

%(slow driving vehicles are known to be more emission intensive)

%\subsubsection{AMoD and Its Potential}
%

%Congestion. Most current private vehicles are designed to support a maximum speed of over 100 miles per hour. However, average urban driving speed is typically around 30 km/hour. 

Extending the concept of MoD to AMoD will not only directly inherit the benefits of autonomous driving technology as mentioned in Section \ref{disruptive_transformation}, but also provide complete solutions to the above limitations of MoD systems. We summarize the promising benefits of AMoD systems as follows:
\begin{itemize}[leftmargin=*]
	\item \textit{Reduction in Number of Vehicles}. An intelligently coordinated AMoD system has the potential to dramatically reduce the number of vehicles needed in a certain area. As reported by a joint study between Stanford and the Singapore-­MIT Alliance for Research and Technology \cite{case_study_AMoD}, if all forms of current transportation in Singapore are replaced by well-designed AMoD systems, only one-third of the current number of vehicles will be needed. 
	
	\item \textit{Reduction in Parking Space}. Parking space is a limited resource in all major metropolises, e.g., the average parking space in Hong Kong costs about 286 thousand USD \cite{parking_cost}. Allocating valuable land to parking in core downtown areas will inevitably increase renting cost and parking costs, which is definitely not aligned with the target of urban sustainability. AMoD systems have the potential to, on the one hand, increase the utilization factor of parking spaces as autonomous vehicles can be stacked behind each other with minimum space, according to a recent study \cite{Nourinejad2018}; on the other hand, reduce the demand of parking since fewer vehicles will be needed and moreover, less parking time will be required since the utilization factor of vehicles can be significantly increased in AMoD systems. 
	
	\item \textit{Autonomous Balancing}. As a fundamental challenge to the operation  of current MoD systems, the problem of being unbalanced yields unnecessary manpower cost and causes extra traffic. Meanwhile, it also influences the service quality as customers will be discouraged by the unavailability of vehicles nearby. AMoD systems have the potential to solve all these issues at its core as vehicles can be autonomously balanced given their being driverless.

	\item \textit{Autonomous Charging}. As we mentioned earlier, the long charging time of EVs is considered to be one of the major obstacles preventing the wider adoption of EVs. AMoD systems have the potential to remove the necessity of owning a private vehicle, which will not only reduce the number of vehicles but also reduce owner waiting times while their EVs are being charged. Moreover, vehicles can autonomously reach preferred charging stations whenever needed. Therefore, AMoD systems have the potential to achieve \textit{anytime}, \textit{anywhere}, and \textit{autonomous} charging without human intervention. In addition to providing AMoD services, a fleet of autonomous EVs is also an excellent source of demand-side flexibility for the implementation of autonomous DSM programs in future energy systems.  We will discuss the details about this opportunity in the next section.
\end{itemize}

In summary, AMoD systems promise to achieve economic, environmental and social sustainability as a whole with autonomous EVs serving as major carriers, augmented with a user-centric and shared mobility based MoD platform, and further empowered with electricity supply from power grids that have high penetrations of DERs and advanced autonomous DSM techniques. It is, therefore, widely anticipated that AMoD systems will be the future of personal mobility in the 21st century \cite{Mitchell2013}.

%\begin{itemize}
%	\item Trip Planning and Booking. 
%	
%	\item Real-time information and feedback. For instance, customers should be able to see the vehicle availability within an app, get the feedback information such as pricing, estimated waiting time and traveling time instantaneously.
%	
%	\item Heterogeneous QoS Requirements.
%	\begin{itemize}
%		\item Shared or Non-shared Mobility.
%		\item Vehicle Type/Size.
%		\item Waiting time and Time of Ride.
%	\end{itemize}	
%\end{itemize}
%
%\textit{Service Providers of AMoD}
%\begin{itemize}
%	\item Real-time Monitoring. 
%	
%	\item Trip Matching (for shared mobility)
%	
%	\item Vehicle-to-Customer Dispatching.
%	
%	\item Vehicle Routing.
%	
%	\item Revenue Management, namely, Pricing.
%	
%	\item Energy refueling, namely charging.
%\end{itemize}
%
%
%\textit{Urban Sustainability}
%

\section{AFoD: An Autonomous DSM \\Program enabled by AMoD Systems}\label{AFoD_overview}
This section first introduces the importance of  flexibility-as-a-service in energy systems and then presents the definition of AFoD based on the classification of different types of flexibility associated with charging autonomous EVs. After that, a conceptual comparison between AMoD and AFod is discussed. Furthermore, we identify four key decisions in creating a synergy between AFoD and AMoD.

\subsection{Motivation: Flexibility-as-a-Service}\label{motivation_flexibility}
According to the International Energy Agency \cite{InternationalEnergyAgency2011}, the definition of \textit{flexibility} of a power system refers to the extent to which a power system can modify electricity production or consumption in response to variability. Another source defines flexibility as follows: \textit{flexibility refers to the capability of modifying the generation injection and/or consumption in reaction to an external signal (e.g., price signal or regulation signal) in order to provide a certain type of service within the energy system} \cite{EURELECTRIC2014}.  Given the fact that the power grid is undergoing an ever-increasing integration of intermittent DERs, our ability to control the power generation to meet the demand is decreasing. \textit{It is therefore of vital importance that flexibility can be elicited from the demand side as a service (namely flexibility-as-a-service) such that a less expensive and more carbon-emission friendly energy systems can be realized. This motivates us to investigate the potential of AFoD}.

\subsection{MDF of EV Charging and Definition of AFoD} 
Flexibility in our context is closer to the definition from \cite{EURELECTRIC2014}, which in general can include the capability of changing the amount of energy to be consumed, the power consumption (further include the rate of change and the response time), the duration, and the location, etc. In particular, we focus on the demand-side flexibility and aim to investigate the different types of flexibility that autonomous EVs can provide to the energy systems. Specifically, charging of autonomous EVs has the flexibility in the following dimensions:
\begin{itemize}[leftmargin=*]
	\item \textit{Power-Flexibility}. EVs can be charged with a variable charging rate. This dimensional flexibility is called \textit{rate-flexibility} or \textit{power-flexibility}. This is the most common flexibility whose benefit has been extensively studied, e.g., ``peak-shaving" in power systems \cite{Wang2013}. 
	
	\item  \textit{Energy-Flexibility}. With the advancement of EV networks, it will be less urgent for an EV to charge as full as possible during every single charging period. Therefore, the energy demand for each charging period can be flexible (e.g., charge more or less during a safe range) as long as there is a proper economic incentive. This dimensional flexibility is called \textit{energy-flexibility}.

	\item \textit{Temporal-Flexibility}. Charging requests of autonomous EVs may not be strict in terms of their charging durations, namely, it is possible to alter the charging period to some extent. This dimensional flexibility is called \textit{temporal-flexibility}.
	
	\item \textit{Spatial-Flexibility}. An autonomous EV is capable of flexibly choosing a preferred charging location, as long as it is  within the EV's reachability. This dimensional flexibility is called \textit{spatial-flexibility}.
\end{itemize}

In summary, multi-dimensional flexibility (MDF) can be derived from the charging of autonomous EVs. MDF is, therefore, an excellent source of demand-side flexibility in energy systems \cite{Bain&Company2018, flexibility_on_demand}. 

%In principle, all the above four-dimensional flexibility exists for traditional EVs or autonomous EVs. 

% We show these multi-dimensional flexibility in Fig. \ref{MDF}. We give examples of the eight cases shown in Fig. \ref{MDF} as follows:
% \begin{itemize}
% 	\item Case 1: This cases corresponds to the scenario when an EV specifies its total energy demand with a chosen charging location and a fixed charging duration. Extensive study has been performed in this case on various aspects such as: impact analysis [], distributed charging coordination [], etc.
% 	
% 	\item Case  
% \end{itemize}

%\begin{figure}
%	\centering
%	\includegraphics[width=8cm]{MDF}
%	\caption{Illustration of the flexibility in energy-, temporal- and spatial-domain.}
%	\label{MDF}
%\end{figure}

%If we fixed the energy, deadline and maximum charging rate, the only flexibility of EV charging is the deferrable property. This is the traditional problem setting for most of the current research. However, if we further consider the energy-flexibility, the deadline-flexibility and the rate-flexibility, then it is possible to exploit the \textit{multi-dimensional flexibility} (MDF) of elastic load to improve the state-of-the-art. 

Motivated by the vital importance of flexibility-as-a-service in the demand side, given the fact that charging of autonomous EVs is an excellent source of MDF, we draw the same analogy as mobility-as-a-service in AMoD and define AFoD as follows: 
\begin{definition}[\textbf{AFoD}]
	Autonomous flexibility-on-demand, or AFoD for short, is a type of energy service where customers can access multi-dimensional flexibility from autonomous EVs as a service on demand. As the service provider, EVs will autonomously drive to the desired location to perform the requested charging and/or discharging as a service.
\end{definition}
Similar to mobility-as-a-service in AMoD, the idea behind AFoD is that different
dimensions of flexibility can be nested using the notion of a service. \textit{Conceptually, AFoD can be regarded as a specific example of an autonomous DSM program. However, given the key idea of leveraging existing infrastructures of AMoD systems, AFoD introduces a considerable amount of unique challenges compared to the broader techniques of autonomous DSM}. The detailed discussion of research challenges is delayed to Section \ref{Challenges}.

\begin{figure}
	\centering
	\includegraphics[width=10cm]{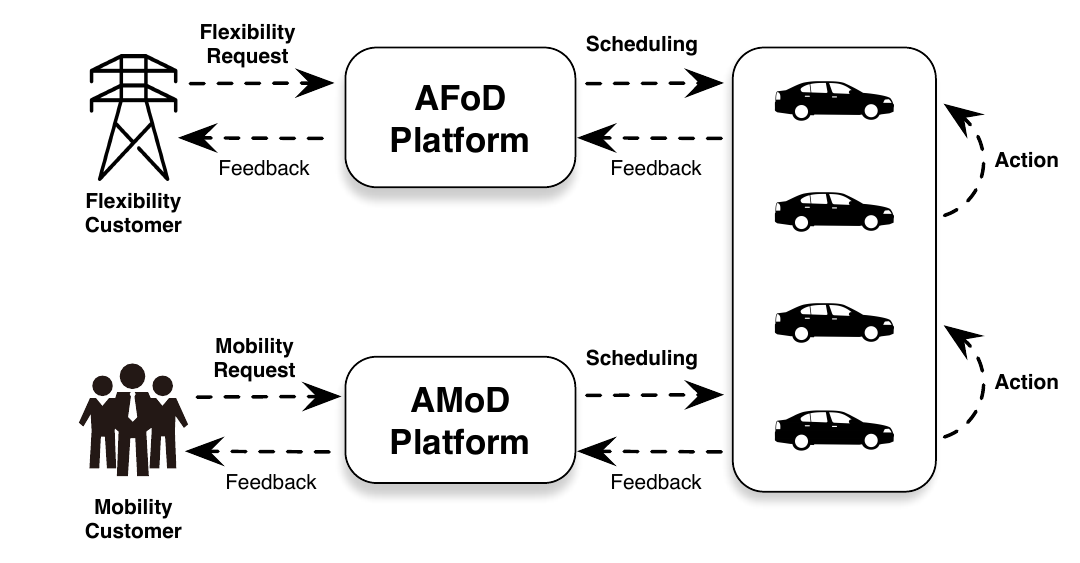}
	\caption{Conceptual comparison between AMoD and AFoD services.}
	\label{Comparison}
\end{figure}

\subsection{Comparison between  AFoD and AMoD}\label{Comparison_AMoD_AFoD}
 
Fig. \ref{Comparison} compares the implementation of AMoD and AFoD. For both of these two services, the systems consist of the following components: 
\begin{itemize}[leftmargin=*]
	\item \textit{Customer}. Customers in AFoD systems can be  power grid operators or DERs with excess energy, e.g., buildings with rooftop solar, while customers in AMoD are human or goods. Note that the number of  AFoD customers can be limited, and in some cases, it can be a single entity, e.g., a single power grid operator who periodically sends flexibility requests to ask for frequency regulation \cite{Kempton2008}. In comparison, the number of AMoD customers is usually enormous and typically exhibits spatial-temporal uncertainty.
	  
	\item \textit{Request}. Depending on different application scenarios, flexibility requests can have different forms. For instance, a building with excess solar power can request a certain number of EVs with almost empty batteries to consume its excess energy during a certain duration, while in providing ``peak-shaving" service to the power grid, the customer may request a certain number of EVs with full batteries to discharge during a certain duration at some locations (this is known as vehicle-to-grid technologies in power systems \cite{Wang2013}.). In comparison, mobility requests usually consist of origin, destination, and number of travelers, etc.
	
	\item \textit{Platform}. The platforms for both services are similar. They both can be a real entity such as a profit-oriented company (e.g., Uber and DiDi) or a nonprofit-oriented (neutral) independent system operator. The two platforms could also converge in future smart cities. Meanwhile, it is even possible to implement AMoD and AFoD in a fully decentralized manner without platforms in the middle. 
	
	\item \textit{Schedule}. The schedule of both platforms contains signals of dispatching a certain vehicle to a certain location. In practice, the two platforms will have different objectives to schedule their respective optimal decisions. We will elaborate more in the next bullet.
	
	\item \textit{Action}. The vehicles' actions to perform AFoD services are essentially different from those to carry out AMoD services. In particular, the AFoD platform schedules vehicles to perform the required charging/discharging service at some designated locations, while the AMoD platform schedules vehicles to pick-up the mobility customer (human or goods) at a designated location (origin) and then deliver the customer to another designated location (destination). Therefore, vehicles to perform AFoD service basically need to keep \textit{idle} (i.e., park and plug-in) at the required location for a certain duration to perform the charging and/or discharging services. In comparison, vehicles should keep \textit{moving} to carry out the AMoD service. Therefore, from this perspective, the vehicles' actions  of AFoD and AMoD are in contention to each other. 
\end{itemize}

In summary, though many similarities can be found between AMoD and AFoD, these two services have essentially different scheduling objectives and vehicles' actions. It is thus very important to investigate how to coordinate these two services in future smart cities. In the following subsection, we will point out the key decisions that, if appropriately coordinated, can lead to a synergy between AMoD and AFoD.

\begin{figure}
	\centering
	\includegraphics[width=7cm]{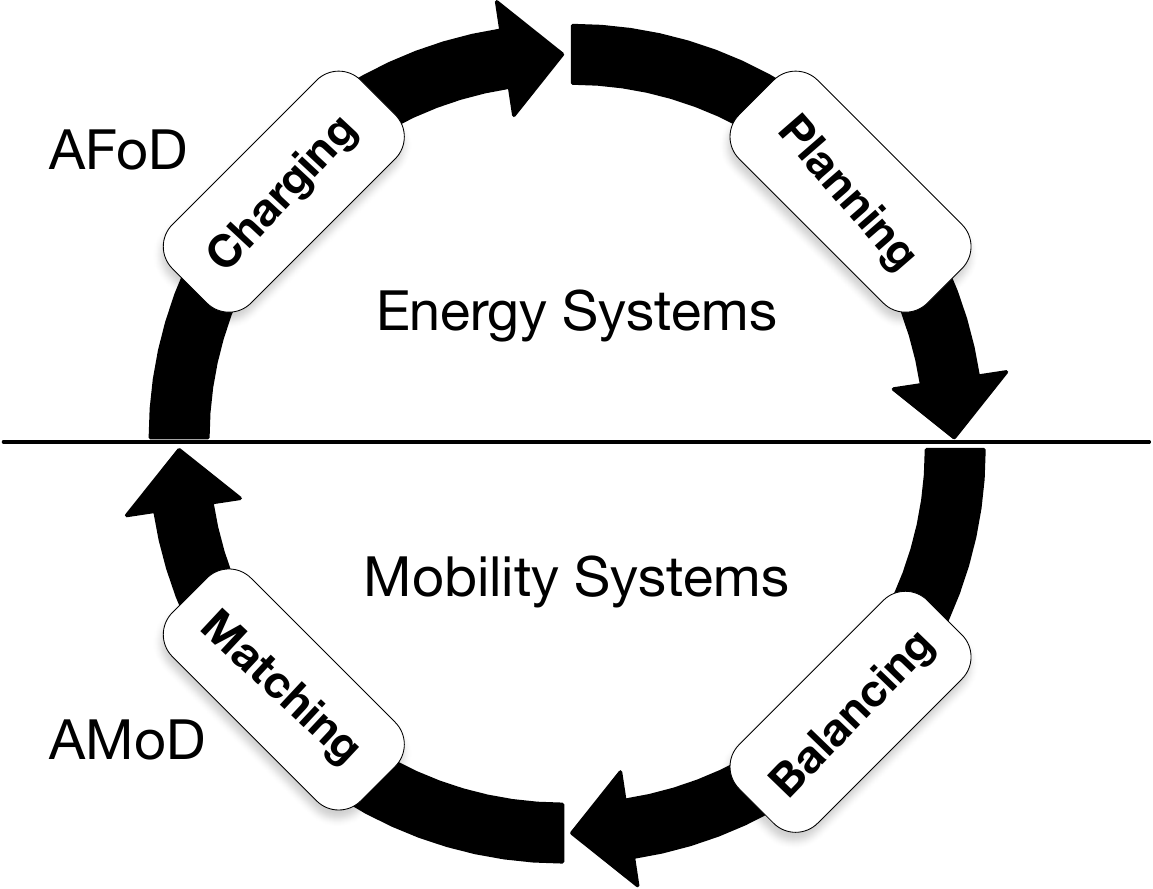}
	\caption{Key decisions to coordinate  AMoD and AFoD  in future smart cities. Circled arrows mean that each decision will be influenced by its previous one.}
	\label{synergy}
\end{figure}

\subsection{Key Decisions to Coordinate AFoD and AMoD}\label{Synergy_AFoD_AMoD}
In this subsection, we classify the following four key decisions, as shown in Fig. \ref{synergy}, that if wisely coordinated with a system-wide outlook, can create a synergy between AFoD and AMoD. 
The four decisions are as follows: i) \textit{planning of infrastructures in energy systems, ii) balancing of vehicles in mobility systems, iii) matching between customers and vehicles (i.e., vehicle-to-customer dispatching) and matching between paths and vehicles (i.e., origin-to-destination routing), and iv) charging control of EVs}. 

%Note that the two-layered view in Fig. \ref{synergy} is a simplification of the three-layered architecture shown in Fig. \ref{AMoD}.  

We show these four types of decisions in Fig. \ref{synergy} with circled arrows, where all decisions can influence the next one they point to. Below we explain in detail about how each of these four decisions can positively contribute to one they point to\footnote{We organize our presentation according to the timescales of each decision, namely from long-timescale infrastructure planning to short-timescale charging scheduling. }.
\begin{itemize}[leftmargin=*]
\item \textbf{Planning}. The first important decision is the infrastructure planning in energy systems, which usually occurs in the timescale of several years or even decades long. As suggested by \cite{Bain&Company2018}, current charging infrastructure planning should focus on reducing the ``range-anxiety" issue, and prioritize charging hubs with grid edge technologies such as integration of DERs and autonomous DSM. A carefully-designed planning decision can not only reduce the total carbon emission from the power sector but also lead to total traveling distance reduction in mobility systems. As a result, the reachability of charging stations will be increased and thus contribute to service quality enhancement in AMoD systems. Therefore, planning decisions can positively contribute to balancing decisions. 

%Infrastructure planning is usually capital intensive and difficult to alter once deployment. Therefore, it is critical to anticipate future trends, challenge the robustness of current business models and question whether future evolutions are being correctly foreseen.

\item \textbf{Balancing}. Autonomous balancing of vehicles in AMoD systems can be performed, e.g., daily or hourly, depending on the periodic patterns of the provided mobility services. As mentioned earlier, balancing is important to guarantee the service quality of AMoD systems. Balancing decisions in AMoD systems will be mainly influenced by the spatial-temporal distribution of mobility requests, but also the planning decisions of charging infrastructures as we mentioned above. A well balanced AMoD system can directly lead to total traveling time reduction since less distance will be traveled to pick-up customers. Therefore, balancing decisions can positively contribute to matching decisions.

%but also be aware of the network of charging stations.

%One the one hand, if the network of charging facilities is designed with consideration of spatial-temporal distribution of mobility requests, we can minimize the minimum extra traffic of balancing will be required. From another perspective,  w

%Leverage the mobility-on-demand business model, together with ride-sharing and clean energy supply, to reduce pollution and congestion, etc. However, the key operational challenge for AMoD systems is \textit{imbalance}: vehicles will naturally concentrate in a subset of areas, limiting the availability in other regions.

\item \textbf{Matching}. Matching in AMoD systems will be performed on demand, i.e., each mobility request will trigger the decision-making\footnote{In theory, matching decisions are performed in real-time, but less frequently to change than charging of EVs. For example, after dispatching a vehicle to pick-up a customer, the vehicle runs a routing algorithm to autonomously navigate and determine the best route. After that, it is unlikely for the vehicle to frequently change its route in the middle of the traveling.}. If appropriately designed, the matching algorithms should be able to dispatch not only the nearest vehicle to the requested customer but also the vehicle with a battery that has an appropriate state-of-charge, since it may be possible to carry out some AFoD services near the destination of current mobility customer. Therefore, a well-designed matching algorithm should not only finish the AMoD service with a performance guarantee but also contribute to the value enhancement of charging control of autonomous EVs.

%Less congested traffic gives vehicles higher charging flexibility since less time is wasted on traveling and idling. 

\item \textbf{Charging}. Charging of EVs will be performed in real-time, depending on the specific external signals as we mentioned in Section \ref{motivation_flexibility}.  For example, if the target of AFoD service is to smooth out the electricity generated by a wind farm, then the charging algorithm should be able to adjust the charging/discharging rates of autonomous EVs according to the fluctuation of wind speed. Charging decisions will also show influence to the design of charging facilities. For instance, a well-coordinated charging control of EVs can guarantee that the total power consumption is within the capacity limit of power distribution network transformers \cite{sunbo}, which thus positively contribute to the planning of infrastructures in energy systems.
 
%vehicles to be charged at a fixed rate and  variable rates requires different charging facilities. 

\end{itemize}

%\begin{enumerate}
%	\item Leverage the flexibility to absorb uncertainty from  distributed energy resources in power networks. 
%	\item Better power supply in terms of sustainability and availability, which thus eases the “range-anxiety” issue and increase the level of mobility. 
%	\item Leverage the mobility-on-demand business model, together with ride-sharing and clean energy supply, to reduce pollution and congestion, etc. 
%	\item Less congested traffic gives vehicles higher charging flexibility since less time is wasted on traveling and idling. 
%\end{enumerate}

%We envision that the synergy effect between AMoD and AFoD.
Based on the above analysis, it is clear that if these four decisions are appropriately coordinated, each of them will positively contribute to the next one, and thus a synergy effect between AMoD and AFoD can be created. Therefore, although the concept of AMoD and AFoD is still in its infancy, all the related stakeholders are encouraged not only to take actions based on current mobility patterns and vehicle ownership models (infrastructure planning in particular as it is in long-timescale), but also anticipate the transformation of mobility and energy in the near future (e.g., AMoD and AFoD services) and thus take actions beforehand accordingly.

\section{Challenges}\label{Challenges}
Despite the immense opportunities of AMoD and AFoD services, especially the appealing synergy effect between AFoD and AMoD in improving urban sustainability,  there exist multiple challenges that must be tackled before the full potential of AMoD and AFoD can be realized in future smart cities. In this section, we provide a short discussion on some of these key challenges.

\subsection{Core Technology Limitations}
The future of AMoD and AFoD relies on multiple core technologies that are currently flawed. 
\begin{itemize}[leftmargin=*]
	\item First, despite the rapid progress of autonomous driving technologies over the past few years,  there is still a long way to go before autonomous vehicles are ready for mass market availability \cite{BobMcDonald}. 
	
	\item Second, although we have seen consistent increases in
	performance and reductions in price of lithium-ion batteries, current battery technologies are in general not cost-effective for implementation of AFoD services. The limitation of current battery technologies also affects the integration of renewables in power grids. For example,  according to  \cite{wind_curtailed}, 39\% of wind energy is being curtailed in Gansu Province, China since the power grid lack the storage systems that can digest this much non-dispatchable wind power.

\end{itemize}

%One the one hand, the   Without a clean generation of electricity, it is impossible to create a sustainable urban future even the autonomous driving technology becomes mutual in the coming decade. 

Some other technologies also play a critical role in shaping the future of AMoD and AFoD. For example,  5G communications, artificial intelligence (e.g., machine learning and big data techniques). However, given the rapid progress and active research of these technical areas, it is unlikely for them to be the major obstacles preventing the success of AMoD and AFoD in future smart cities.

\subsection{Regulation and Public Acceptance}
Although the core technologies of autonomous driving and batteries are flawed, it is widely believed that these technologies themselves might not be the biggest hurdle to the success of AMoD  \cite{disruptive_technologies}. Rather, it is the regulation that is preeminent. Many legal and ethical questions remain to be addressed, and some of them are quite problematic. For example, who should be responsible for an accident caused by autonomous vehicles; how to program the autonomous vehicles to make decisions under  situations involving a dilemma, etc.  Meanwhile, gaining public support is also another big challenge. After all, the realization of the full potential of AMoD systems is inherently tied to the residents' willingness to adopt such kind of technologies and business models.

\subsection{Challenges for System-wide Coordination}
In addition to the core technology limitations and the regulation problems, there also exist some key questions that require a deeper investigation, and all of these open questions are highly related to the performance of system-wide coordination. For instance, 
\begin{itemize}[leftmargin=*]
	\item Will AMoD really contribute to traffic and congestion reduction? This is a key criticism for existing MoD systems and future AMoD systems \cite{Gehrke2018, increased_congestion}. Indeed, on the one hand, AMoD systems will lead to a reduction of vehicle ownership, but on the other hand, many autonomous vehicles may run on the road before picking up their next customers. Meanwhile, the necessity of balancing also creates extra traffic. Therefore, it is unclear whether the development of AMoD systems will really contribute to traffic and congestion reduction or not. 
	
	\item Will AFoD really be cost-effective, and how to coordinate AFoD and AMoD in real-time? The key assumption of AFoD is that autonomous EVs can provide a low-cost and carbon-free alternative to help the integration of DERs in future energy systems. However, there exist quite a few questions. For instance, would it be better for energy systems to use dedicated batteries instead of autonomous EVs? If AFoD is really cost-effective, how significant when compared to using dedicated batteries? Meanwhile, having an optimal real-time coordination between AMoD and AFoD is very challenging given the dynamics of urban mobility and energy systems.
\end{itemize}

It is worth emphasizing that the answers to the questions above highly rely on whether there is an effective system-wide coordination, which further relies on if we have a deeper understanding of the following key research challenges:
%In fact, it is essential to fully realize the potential of AMoD and AFoD. 
% Below we summarize some of the key research directions that are critical. 

%However, given the multifaceted nature of AMoD systems, many problems remain unclear and require an in-depth research.

\subsubsection{High-dimensional Systematic Uncertainties} Coordination of AMoD and AFoD in future mobility and energy systems must address high-dimensional uncertainties from multiple aspects, including weather, traffic, and renewable energy generation, etc. The high-dimensional uncertainties will bring considerable challenges to the following directions:
\begin{itemize}[leftmargin=*]
	\item \textit{Real-time Monitoring}. Real-time monitoring is critical to enable quick reactions to momentary events in AMoD and AFoD systems.  It is in some sense the first and most important step before implementing any type of system-wide coordination algorithms. For instance, the CVST project from the University of Toronto is motivated by this and aims to provide a real-time monitoring of transportation networks of Toronto \cite{cvst}.
	
	\item \textit{Data-Driven Prediction}. Predication will play an important role in key decisions such as balancing and matching. Some recent studies such as \cite{Lu2013} show that it is promising to predict mobility patterns based on mobile phone call data. However, the coordination of AMoD and AFoD relies not only on the prediction of human mobility  patterns, but also the prediction of many factors such as renewables in the power grid.
	
%	\item \textit{Robustness of Coordination}. The high-dimensional uncertainties also lead to challenges to ensure the robustness of coordination. 
\end{itemize}

Both real-time monitoring and prediction are critical for the coordination between AMoD and AFoD. Further research is needed along these two directions in order to create a synergy between AMoD and AFoD.

%In order to have a system-wide coordination, it is expected that both real-time monitoring and accurate data-driven prediction should be used.

\subsubsection{Complexity of Key Operational Algorithms} Another important research direction is on the key algorithms to support the system-wide coordination between AMoD and AFoD. For example, \textit{dial-and-ride problems} \cite{Cordeau2007} are essential to decisions such as matching and balancing. However, dial-and-ride problems are historically known to be NP-hard and are considered to be very complex problems. Moreover, \textit{restricted (stochastic) shortest path problems} \cite{Hassin1992} are another type of classic NP-hard problems that are critical for balancing, matching, and charging of autonomous EVs in the coordination of  AMoD and AFoD. 

\subsubsection{Economic Incentive and Mechanism Design}
The three-layered architecture shows that future urban mobility and energy systems consist of multiple self-interested agents and groups. Ensuring collaboration among different urban infrastructures is considered to be one of the key factors in the development of future smart cities  \cite{mechanism_smart_cities}. However, without appropriate business models, economic incentives and mechanisms, it will be very difficult if not impossible to directly coordinate such a complex system, even the above two challenges are successfully solved.
%For example, sources have suggested that even in transition advanced nations such as Germany, the full potential of residual load flexibility has still not been performed [28][29]. 

In summary, research into AMoD and AFoD, especially the coordination between AMoD and AFoD, are still in its infancy. Although both of these two services have a great potential in creating urban sustainability in future smart cities,  further research into the system-wide coordination is needed to fully realize their potential.

%although there exist some core technology limitations and lack of a clear regulatory framework and public acceptance, public- and private-sector stakeholders are encouraged not only to take actions, infrastructure planning in particular, based on current patterns of mobility and vehicle ownership model, but also anticipate the transformation of mobility and energy in the near future and thus take action beforehand accordingly \cite{Bain&Company2018}.

\section{Conclusions}\label{Conclusions}
In this paper, we gave an overview of the management of two promising autonomous mobility and energy services in future smart cities, namely the AMoD and AFoD services. We proposed a three-layer architecture to show the convergence of future mobility and energy systems.  For each layer, we provided a brief overview of the disruptive transformations that directly contribute to the rise of AMoD systems. Next, we proposed a new concept called AFoD, which is a promising energy service platform built directly on top of AMoD systems. In the vision of AFoD, autonomous EVs can provide their charging flexibilities as a service in energy systems by performing autonomous charging/discharging services on demand. In addition, we performed a conceptual comparison between AMoD and AFoD, and further identified four key decisions that, if appropriately coordinated, will create a synergy between AMoD and AFoD. Finally, we briefly summarized some key challenges towards the success of AMoD and AFoD in future smart cities and presented some open research questions regarding the coordination between AMoD and AFoD.

%\section*{Acknowledgment}
%(Please let me know if we need to Ack some fundings here). 
%The work is supported by ... 

\bibliography{mybib}{}
\bibliographystyle{IEEEtran}

\end{document}